\documentclass{aa}  
\usepackage{hyperref}
\usepackage{natbib}
\bibpunct{(}{)}{;}{a}{}{,}
\usepackage{amsmath, amsfonts, amssymb} 
\usepackage{wasysym}
\usepackage{siunitx}
\usepackage{paralist}
\usepackage[center]{caption}
\usepackage{placeins}
\usepackage{xcolor}
\hypersetup{
    colorlinks,
    linkcolor={red!50!black},
    citecolor={blue!50!black},
    urlcolor={blue!80!black}
}

\usepackage{graphicx}
\usepackage{comment}
\usepackage{txfonts}
\usepackage[normalem]{ulem}

\begin{document} 

   \title{Temperature and differential emission measure evolution of a limb flare on 13 January 2015}

   \titlerunning{Temperature and DEM evolution of a limb flare}

   \author{M. Br\"ose
          \inst{1}\inst{2}\inst{3}
          \and
          A. Warmuth\inst{1}
          \and
          T. Sakao\inst{4}\inst{5}
                        \and            
                Y. Su\inst{6}\inst{7}                     
          }

   \institute{Leibniz Institut für Astrophysik Potsdam (AIP), An der Sternwarte 16, 14482 Potsdam, Germany\\
             \email{mbroese@aip.de, awarmuth@aip.de}
         \and
             Department of Physics, Freie Universit\"at Berlin,
               Arnimallee 14, 
14195 Berlin, Germany
                \and
                        Zentrum für Astronomie und Astrophysik, Technische Universit\"at Berlin,  Hardenbergstraße 36, 10623 Berlin, Germany
         \and
                Institute of Space and Astronautical Science (ISAS), Japan Aerospace Exploration Agency (JAXA), 3-1-1                                      Yoshinodai, Chuo-ku, Sagamihara, Kanagawa 252-5210, Japan  
         \and
                Department of Space and Astronautical Science, School of Physical Sciences, SOKENDAI,
3-1-1 Yoshinodai, Chuo-ku, Sagamihara, Kanagawa 252-5210, Japan                  
                \and
                        Key Laboratory of Dark Matter and Space Astronomy, Purple Mountain Observatory, Chinese Academy of Sciences,10 Yuanhua Road, Nanjing 210023, China
                \and
                        School of Astronomy and Space Science, University of Science and Technology of China, 96 Jinzhai Road, Hefei 230026, China                    
             }

  \abstract
  {Spatially unresolved observations show that the cooling phase in solar flares can be much longer than theoretical models predict. It has not yet been determined whether this is also the case for different subregions within the flare structure.}
    {We aim to investigate whether or not the cooling times, which are observed separately in coronal loops and the supra-arcade fan (SAF), are in accordance with the existing cooling models, and whether the temperature and emission measure of supra-arcade downflows (SADs) are different  from their surroundings.}
  {We analysed the M5.6 limb flare on 13 January 2015 using SDO/AIA observations. We applied a differential emission measure (DEM) reconstruction code to derive spatially resolved temperature and emission measure maps, and used the output to investigate the thermal evolution of coronal loops, the SAF, and the SADs.}
  {In the event of 13 January 2015,  the observed cooling times of the loop arcade and the SAF are significantly longer than predicted by the Cargill model, even with suppressed plasma heat conduction. The observed SADs show different temperature characteristics, and in all cases a lower density than their surroundings.} 
  {In the limb flare event studied here, continuous heating likely occurs in both loops and SAF during the gradual flare phase and leads to an extended cooling phase.}


   \maketitle

\section{Introduction}
The thermal evolution of coronal plasma during solar flares is a key issue in the analysis of energy dissipation in the solar corona. After a massive energy input in the impulsive flare phase, coronal plasma cools by radiation and heat conduction (e.g. \citet{2011ASPC..448..441F, 2020A&A...644A.172W}). However, based on spatially unresolved data it was observed that the cooling phase in solar flares can be much longer than expected (\citep{1980sfsl.work..341M}, \citep{Reeves2002}, \citep{Ryan2013}, \citep{2016ApJ...820...14Q}). This might be a result of reduced heat conduction or ongoing heating. However, a heterogeneous thermal evolution across the extended flare structure is conceivable in light of the existence of different subregions.

Therefore, spatially resolved temperature data or reconstructions are needed to unveil the contributions to the overall thermal evolution in solar flares. In addition to that of the coronal loops themselves, the thermal evolution in regions above them is of great interest. Supra-arcade fans (SAFs) are faint irregular regions of hot plasma observed above flare arcades (\citep{1998SoPh..182..179S}, \citep{1999ApJ...519L..93M}, \citep{2000SoPh..195..381M}, \citep{Innes2014}). They belong to the less understood parts in the standard flare model (\citep{1964NASSP..50..451C}, \citep{1966ApJ...143....3S}, \citep{1974SoPh...34..323H}, \citep{1976SoPh...50...85K}), but are associated with magnetic reconnection and current sheets (\citet{2010AAS...21640423S}, \citet{Liu2013}, \citet{Innes2014}). In particular, their thickness, magnetic configuration, and the origin of their fine structure are not well understood. Part of their fine structure are the supra-arcade downflows (SADs), which are dark moving structures within the SAF \citep{1999ApJ...519L..93M} that are seen as a consequence of the reconnection process in the high corona. Previous studies found them to be less dense and predominantly cooler than the ambient SAF plasma (\citet{1999ApJ...519L..93M, Savage.2012, 2014ApJ...786...95H, 2017ApJ...836...55R}).  Analysing the thermal characteristics of the SAF and SADs may help to link them to the reconnection process in solar flares.

In this paper, these issues concerning the temperature and differential emission measure evolution are addressed with a case study of a limb flare on 13 January 2015. By deriving spatially resolved temperature and density maps with a DEM method, we investigated how different parts of the flare, namely coronal loops and the SAF, evolve over time. The fortunate viewing geometry allowed to study the cooling times independently in two subregions. For both subregions, we compare the cooling times in the gradual phase of the flare with a theoretical cooling model by \cite{1995ApJ...439.1034C}.

First, we describe the flare observations that are used (Sect. \ref{sec: Observations}). Then the DEM reconstruction method (Sect. \ref{sec: DEM Reconstruction Technique}) is introduced and applied to the Atmospheric Imaging Assembly (AIA) data. The obtained DEM outputs (emission measure and temperature maps) are further analysed along linear cuts (Sect. \ref{sec: SAD Observations}) and in certain subregions (Sect. \ref{sec: Thermal Evolution in Subregions}). SADs are studied with respect to their density and temperature in order to illuminate their contribution to the  thermal evolution of the flare  and the results are compared with findings from previous studies. After a brief introduction to the Cargill model, the cooling times that we find for loops and SAF are finally compared with its predictions (Sect. \ref{sec: Cooling Rates of Loops and SAF}).

\section{Observations}
\label{sec: Observations}

In this section, we introduce the flare observations that we used. Geostationary Operational Environmental Satellite (GOES) observations provide an overview of the event, while the main source of information is AIA, because of its emphasis on the spatial evolution of different flare parts. The section is closed by the determination of characteristic lengths of the coronal loops.

\subsection{GOES X-ray flux}
\begin{figure}[h]
\centering
\includegraphics[width=0.95\linewidth]{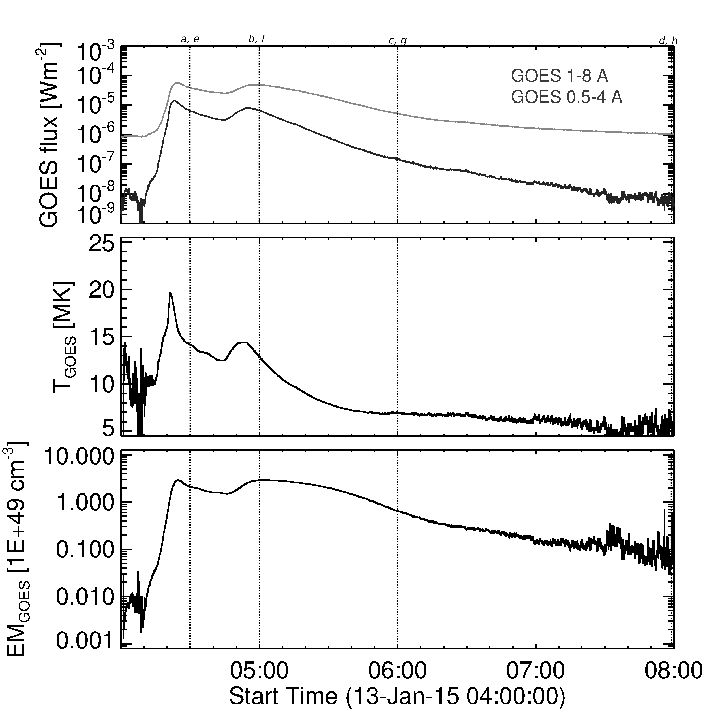}
\caption{GOES observations of the M5.6 flare of 13 Jan 2015. X-ray fluxes (\textit{top}) show two flare peaks, at 04:20 UT and at 04:55 UT. Based on the filter-ratio method, the temperature and EM evolution is plotted below. Vertical dashed lines show the times of the image series in Figs. \ref{fig: mosaic AIA} and \ref{fig: mosaic DEM}.}
\label{fig: GOES_curves}
\end{figure}

The M5.6 flare of 13 January 2015  is characterised by two peaks in the GOES X-ray fluxes (Fig. \ref{fig: GOES_curves}). A first impulsive peak at 04:24 UT is followed by a decrease (first gradual phase) and second peak at 04:58 UT. From there the flux decays towards the initial values (second gradual phase).
\newpage
Assuming an isothermal plasma, a filter-ratio method can be applied to the optically thin emission recorded by the two GOES channels \citep{1973SoPh...32...81V,2005SoPh..227..231W} yielding temperature and emission measure (EM), which are plotted below. The estimated EM and temperatures are based on GOES/XRS full-disk spatially integrated observations. Both background subtracted curves have a double peak shape. The temperature reaches $20~\si{MK}$ and $15~\si{MK}$ at the second peak. These curves already show that the flare is accompanied by a strong heating process, which increases the temperature in a very short time period of the impulsive flare phases. The extended gradual phase is also particularly remarkable, as it lasts for more than three hours.

\subsection{AIA data}
\label{sec: AIA Data}

\begin{figure*}[h]
\centering
\includegraphics[width=0.95\linewidth]{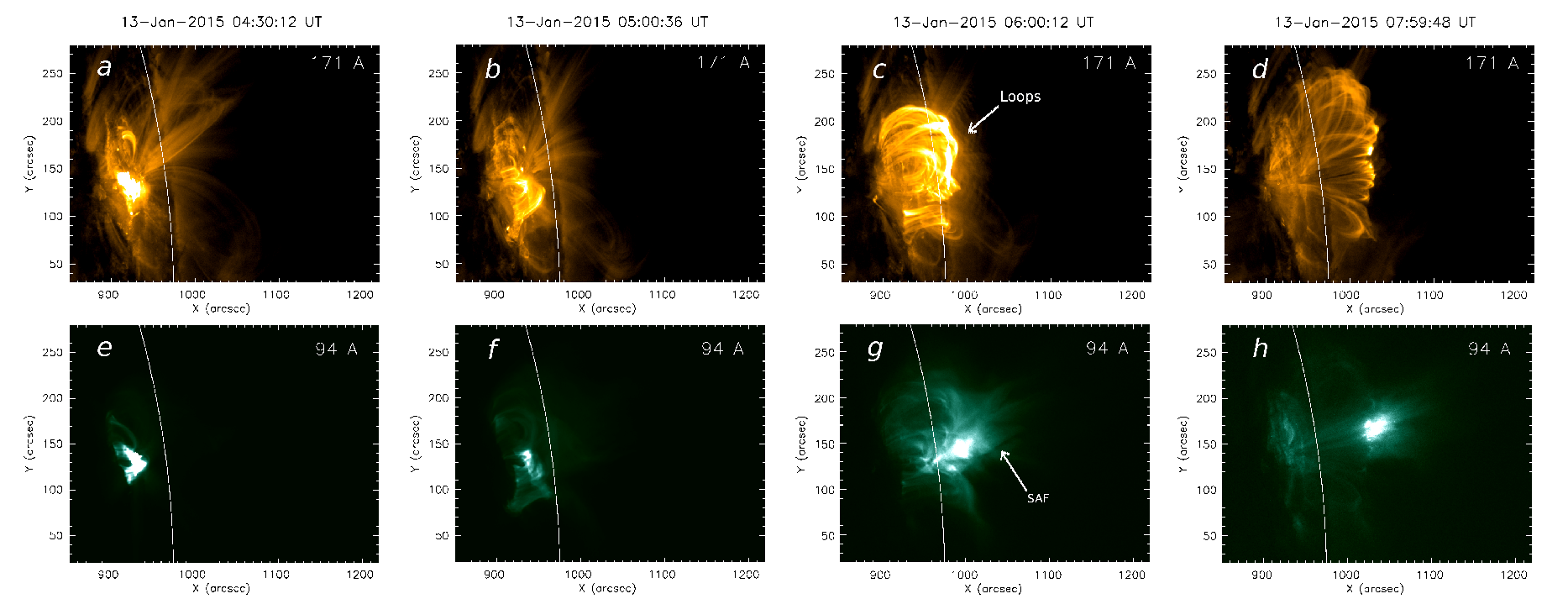}
\caption{Evolution of the flare as observed by AIA.  The growing loop arcade is best seen in the cool $171 \AA$ channel (\textit{top}), while the hotter  $94\AA$ channel (\textit{bottom}) additionally shows a SAF above the arcade.}
\label{fig: mosaic AIA}
\end{figure*}

Spatially resolved observations are needed to reveal the flare structure responsible for this heating and show how the plasma located above the active region reacts to changing magnetic field configuration in the flare process. We use AIA \citep{2012SoPh..275...17L} on board NASA's Solar Dynamics Observatory (SDO), which is a full-disk extreme ultraviolet (EUV) imager with 1.5 arcsec spatial resolution and 12 s temporal resolution. It takes images in seven EUV band passes. 

Figure \ref{fig: mosaic AIA} shows the flare evolution in one hot ($94~\si{\AA}$, sensitivity peaks: $1.1~\si{MK}$ and $7.9~\si{MK}$) and one cool  AIA channel ($171~\si{\AA}$, sensitivity peak: $0.8~\si{MK}$). The first images (Fig. \ref{fig: mosaic AIA}, 04:41 UT) were taken between the two flare peaks, whereas the others show the evolution in the gradual flare phase. The displayed images suggest that the whole flare structure can be  divided  into two distinct parts, namely plasma-filled coronal loops with sharp boundaries and the SAF, a comparatively diffuse region above it. These two regions have different characteristics. The absence of the SAF in the $171~\si{\AA}$ channel already suggests very different temperature structures across the flare.

The evolution starts with the formation of a complex system of entangled loops, which can be best observed in the $171~\si{\AA}$ channel. In the first impulsive and gradual phase of the flare (Fig. \ref{fig: mosaic AIA}, a, e),  there are a few different loop systems visible, each with different loop radii. In the second gradual phase (after the first X-ray peak), the loop structure becomes strongly simplified in comparison  (Fig. \ref{fig: mosaic AIA}, b-c). The loop structure becomes larger, more extended and ordered over the flare period (Fig. \ref{fig: mosaic AIA}, c-d). This is consistent with the expected reduced magnetic complexity due to dissipation of free magnetic energy during the reconnection process.

The region above the loops, that is, the SAF, can be best observed in the channel $94~\si{\AA}$ and is also visible in channel $131~\si{\AA}$. In the beginning of the flare, it starts as a faint curtain and intensifies during the gradual phase. There is a diffuse irregular pattern across the upper boundary of the SAF. Because of its inhomogeneous character and moving spiky shape, it appears much more dynamic than the trapped plasma within the loops.

After 05:00 UT, the SAF decreases in size. In contrast to the loops, it narrows and evolves to a region of higher emission that lies close above the arcades. This core region remains with a significantly higher emission in the $94~\si{\angstrom}$ channel than its surroundings (and even higher than the loops) and remains visible during the entire gradual phase.
In order to extract the full temperature information, which is contained in the intensity data of the multi-thermal AIA channels, a DEM reconstruction code (\citep{2015ApJ...807..143C}, \citep{2018ApJ...856L..17S}) is applied to this data in Sect. \ref{sec: DEM Reconstruction Technique}.

\subsection{Characteristic lengths}
Characteristic lengths are needed for the calculations of cooling times in Sect. \ref{sec: Cooling Rates of Loops and SAF}. Therefore, loop half-lengths and loop heights are measured from AIA images and are listed in Table \ref{tab: Lengths}. The footpoints are used as the base of the loops. The loop half-length is estimated by following the data points. These are distances projected onto the plane of the sky. Due to the position of the  flare at the solar limb, projection effects are assumed to be negligible.

The observed loop system consists of smaller and larger loops, which evolve during the flare process and are therefore measured at three different times. To find their loop half-lengths, channels $171 ~\si{\angstrom}$ (for the cooler loops) and $131 ~\si{\angstrom}$ (for the hotter loops) are used. While observations from channel $94 ~\si{\angstrom}$ revealed the SAF in the greatest detail, channel $131 ~\si{\angstrom}$ is used in this section, because it recorded the hottest and largest loops. Additionally, we measured the SAF and find a width of approximately $50~\si{Mm}$. 

\begin{table}[h]
\centering
\caption{Characteristic lengths in two different AIA channels.}
\begin{tabular}{c|c|c|c}
\hline
\textbf{Wavelength}   & \textbf{Time}         & \textbf{Loop Half-Length} & \textbf{Loop Height} \\
\multicolumn{1}{l|}{} & UT & [$\si{Mm}$]                 & [$\si{Mm}$]            \\ \hline
$131 \si{\angstrom}$                   & 05:00                 & $55$                      & $45$                 \\
                      & 06:00                 & $70$                      & $50$                 \\
                      & 07:00                 & $95$                      & $75$                 \\ \hline
$171 \si{\angstrom}$                   & 05:00                 & $30$                      & $25$                 \\
                      & 06:00                 & $75$                      & $50$                 \\
                      & 07:00                 & $90$                      & $70$           \\ \hline
\end{tabular}
\label{tab: Lengths}
\end{table}


\section{ Reconstruction of the differential emission measure}
\label{sec: DEM Reconstruction Technique}
\begin{figure*}[!]
\centering
\includegraphics[width=0.98\linewidth]{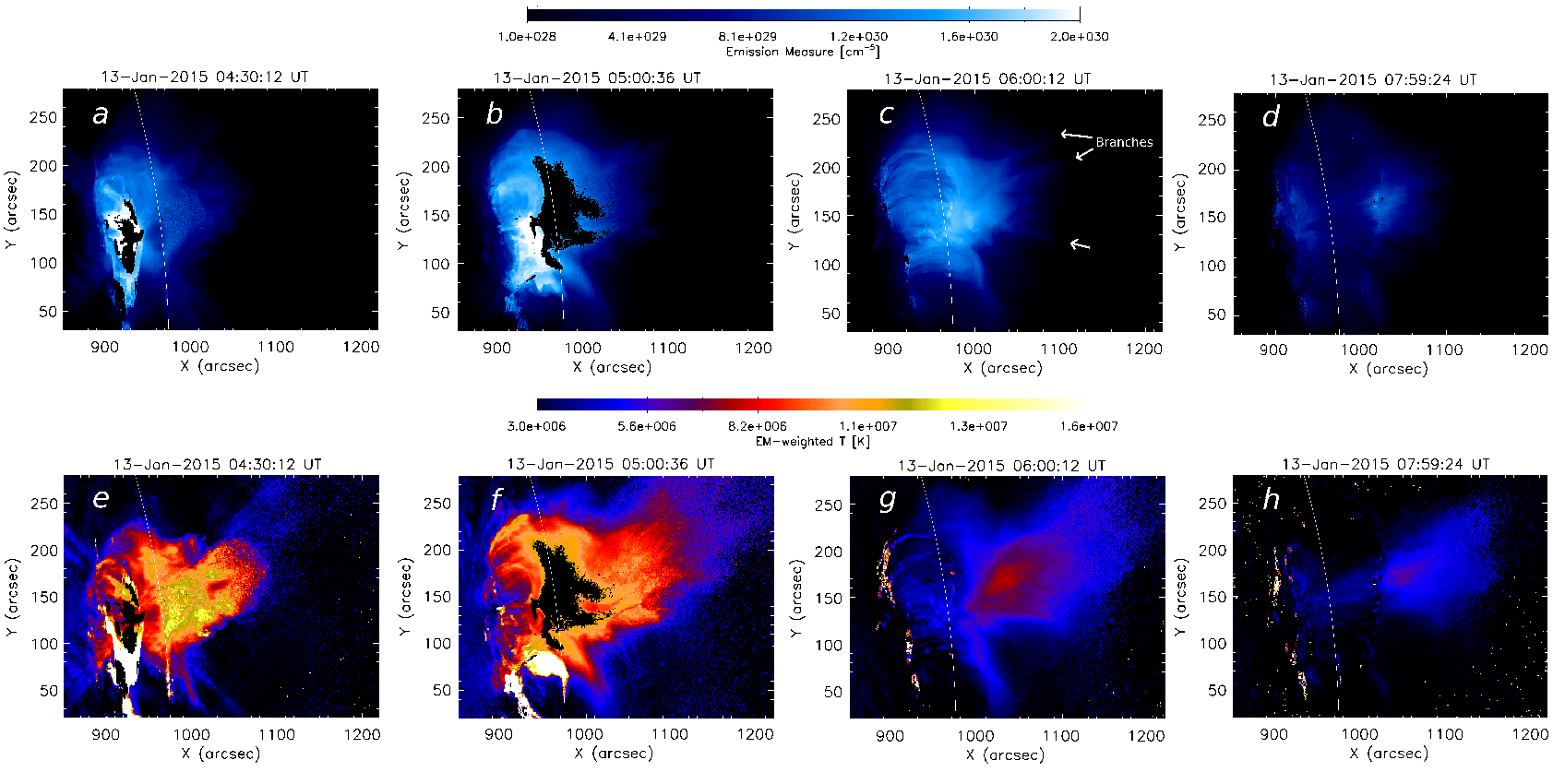}
\caption{Evolution of the flare as seen in total EM (\textit{top}) and EM-weighted temperature (\textit{bottom}). We note the presence of a SAF above the flare loop system, which can be best seen in the temperature maps (\textit{f-h}). White arrows highlight the branches of alternating high and low EM values in the SAF. The black region at the apex of the loop arcade (e.g. panel \textit{b}) is an artefact caused by saturation of one or several AIA channels.}
\label{fig: mosaic DEM}
\end{figure*}

During dynamic heating and cooling processes, plasma contained in structures like coronal loops and the SAF evolves in time. There are different methods to reconstruct temperatures and densities from a set of EUV images. One of them is the DEM reconstruction. The  DEM describes the amount of thermal plasma along the line of sight as a function of temperature $T$ (e.g. \citep{2018ApJ...856L..17S}).

The relation between observed intensities in the different AIA channels $I_i$ and the DEM can be formulated as

\begin{align}
I_i = \int _{0}^{\infty} K_i(T) DEM(T) dT, 
\label{eq: intensities}
\end{align}

where $K_i$ is the temperature response function of the i-th channel. \\

The equations obtained from different EUV channels need to be inverted to find the DEM distribution from the observed intensities. In this paper, the reconstruction is performed with a sparse-inversion DEM code originally developed by \citet{2015ApJ...807..143C}, which was recently modified by \citet{2018ApJ...856L..17S}. In the present paper, we use the latter version of the code because it suppresses the spurious EM contributions at high temperatures which are ignored when several other methods are used (cf. \citet{2018ApJ...856L..17S}).  We verified this by comparing RHESSI \citep{r2002SoPh..210....3L} HXR spectra with the emission synthesised from the derived DEMs, which have shown good agreement at energies where thermal emission dominates (cf. Fig. \ref{fig:RHESSI_DEM_spec2}). Noticeable is that the DEM algorithm of \cite{2018ApJ...856L..17S} provides significantly better agreement with the measured X-ray spectrum, especially at the higher energies.

The code allows us to investigate certain subregions within the flare structure independently from each other (pixel-based method) and to observe how the EM in certain temperature bins evolves over time. DEM and emission measure $EM_T$ within a certain temperature bin $\Delta T$ are related by

\begin{align}
EM_T =& DEM(T) \cdot \Delta T =  \int^{T_1}_{T_0} \int n_e^2(T, z) dzdT, 
\end{align}
where $n_e$ is the electron number density and $z$ is the distance along the line of sight. The temperature range of the DEM analysis spans from $10^{5.5} \si{K}$ to $10^{7.5} \si{K}$, a regime which is adequately covered by the six iron line channels of AIA. The EM is reconstructed for temperature bins with a size of $\log_{10}T$ equal to $0.05$.  


\newpage

EM-weighted temperatures, $T_{EM}$, can be then derived from the DEM distribution by

\begin{align}
T_{EM} = \frac{\int T \cdot DEM(T)dT}{EM},
\end{align}

where $EM$ is the summed EM from all temperature bins. Assuming a certain thickness of plasma along the line of sight allows us to also derive densities from the DEM outputs:

\begin{align}
n_e =\sqrt{\frac{EM}{V}}=\sqrt{\frac{EM}{(A\cdot D),}}
\label{eq: density}
\end{align}

where $V$ is integration volume, $A$ is integration area, and  $D$ is the thickness of the observed structure, that is, loop thickness.

\subsection*{Spatially resolved emission measure and temperature maps}

The pixel-based DEM routine preserves the spatial information of the AIA input data and provides detailed EM and temperature (T) maps. These give greater insight into the physics of the flare process, which cannot be obtained from spatially integrated observations, such as GOES X-ray fluxes.

Figure \ref{fig: mosaic DEM} shows the total EM and EM-weighted T distribution for four time intervals. In the flare process, larger amounts of plasma are quickly heated and become visible in the EM maps. During the impulsive phase, the loop system is initially small and disordered (Fig. \ref{fig: mosaic DEM}a, 04:22 UT) and expands quickly. In this expansion phase the SAF has temperatures of $10~\si{MK}$ in the core and $4~\si{MK}$ at its boundaries. Around the second flare peak at 04:58 UT, the SAF reaches its maximum size. While the loops account for most of the emission measure, the temperatures are rather similar in these two regions (Fig. \ref{fig: mosaic DEM}f). At this stage, both EM and T maps begin to show branches of alternating high and low values at the upper boundary of the SAF (cf. Fig \ref{fig: mosaic DEM}b, \ref{fig: mosaic DEM}f). 

Thereafter, in the gradual phase, loops and SAF begin to cool. A faster temperature reduction in the loops makes the SAF more dominant in the T maps over time (Fig. \ref{fig: mosaic DEM}g). At 06:00 UT, which is one hour after the second flare peak, the highest temperatures can be found close above the flare loops in the SAF core. In general, EM and temperatures are higher in regions located closer to this core region and decrease towards its boundaries. In contrast, the loops show a rather uniform temperature distribution. At the upper boundary of the SAF, alternating low and high EM values are most prominent at that time (Fig. \ref{fig: mosaic DEM}c). These are accompanied by SADs, which are analysed separately in Sect. \ref{sec: SAD Observations}.

Towards the end of the gradual phase, the loops have cooled down almost completely, while the SAF remains hot  with a core temperature of about $6~\si{MK}$ three hours after the impulsive flare phase (Fig. \ref{fig: mosaic DEM}h). Overall, its width has decreased strongly in comparison with earlier times, but it has retained a higher EM in its core than the loops (Fig. \ref{fig: mosaic DEM}d).

In some cases, the DEM code is not able to constrain the DEM. Very low signal or saturation in the AIA input data are possible reasons for spurious DEM results. Saturation occurs for example in the core of the loop region at the flare peak (Fig. \ref{fig: mosaic DEM}, a and e). These pixels are excluded from the analysis and are displayed in black (Fig. \ref{fig: mosaic DEM}, b and f). `Hot pixels' occur in areas with very low EM (e.g. Fig. \ref{fig: mosaic DEM}e). Because the used method is pixel independent, the other parts of the image are not affected and can still be used for the analysis.
\FloatBarrier
\newpage
\section{SAD observations}
\label{sec: SAD Observations}
\begin{figure}[h]
\centering
\includegraphics[width=0.75\linewidth]{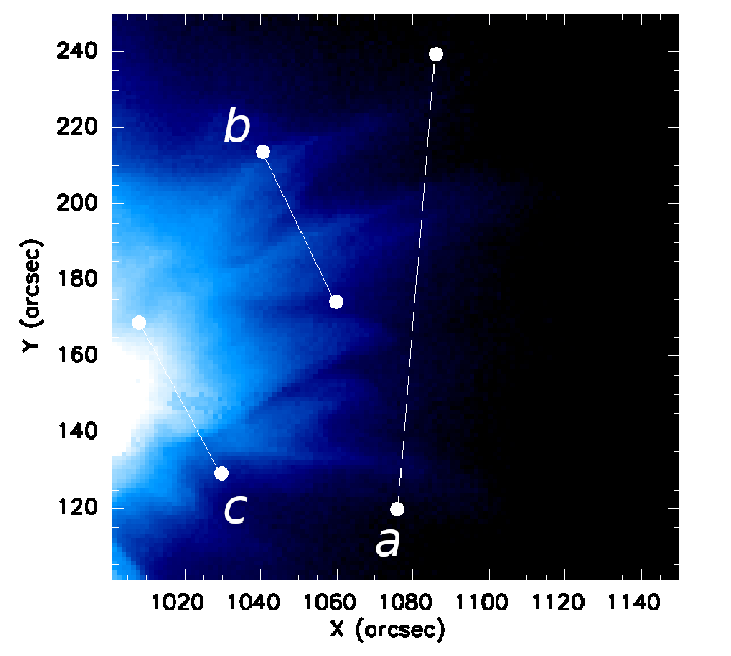}
\caption{Linear cuts \textit{a}, \textit{b,} and \textit{c} for the SAD analysis on the EM map at 05:56:12 UT.}
\label{fig: CuttingLines}
\end{figure}

In the upper part of the SAF, linear cuts (Fig. \ref{fig: CuttingLines}) are used to analyse SADs and the branched patterns in the EM maps. For each time-step, the EM and T values along these lines are extracted and displayed in 2D plots (Fig. \ref{fig: uppercut}). This is done both for a longer `cut-a' through the entire SAF (Fig. \ref{fig: uppercut}) and then also on smaller scales (`cut-b', `cut-c') at positions where SADs are observed. SADs are then identified by a sudden darkening, that is, a  decrease in EM, at a certain distance and time on the linear cut. At these positions, the EM and T are recorded over time (e.g. Fig. \ref{fig: demseriesSAD_comparison_50_62_auschnitt}). Five SADs are studied in this paper  by comparing their EM and T values before, during, and after the passing SAD (Table \ref{tab: SAD}). Additional image sequences of AIA observations, EM, and T maps are used to verify whether observed EM depletions are indeed moving dark structures within the SAF, in the same was as presented in previous studies of SADs.

\subsection{Evolution at the SAF branches}

\begin{figure}[!]
\centering
\includegraphics[width=0.89\linewidth]{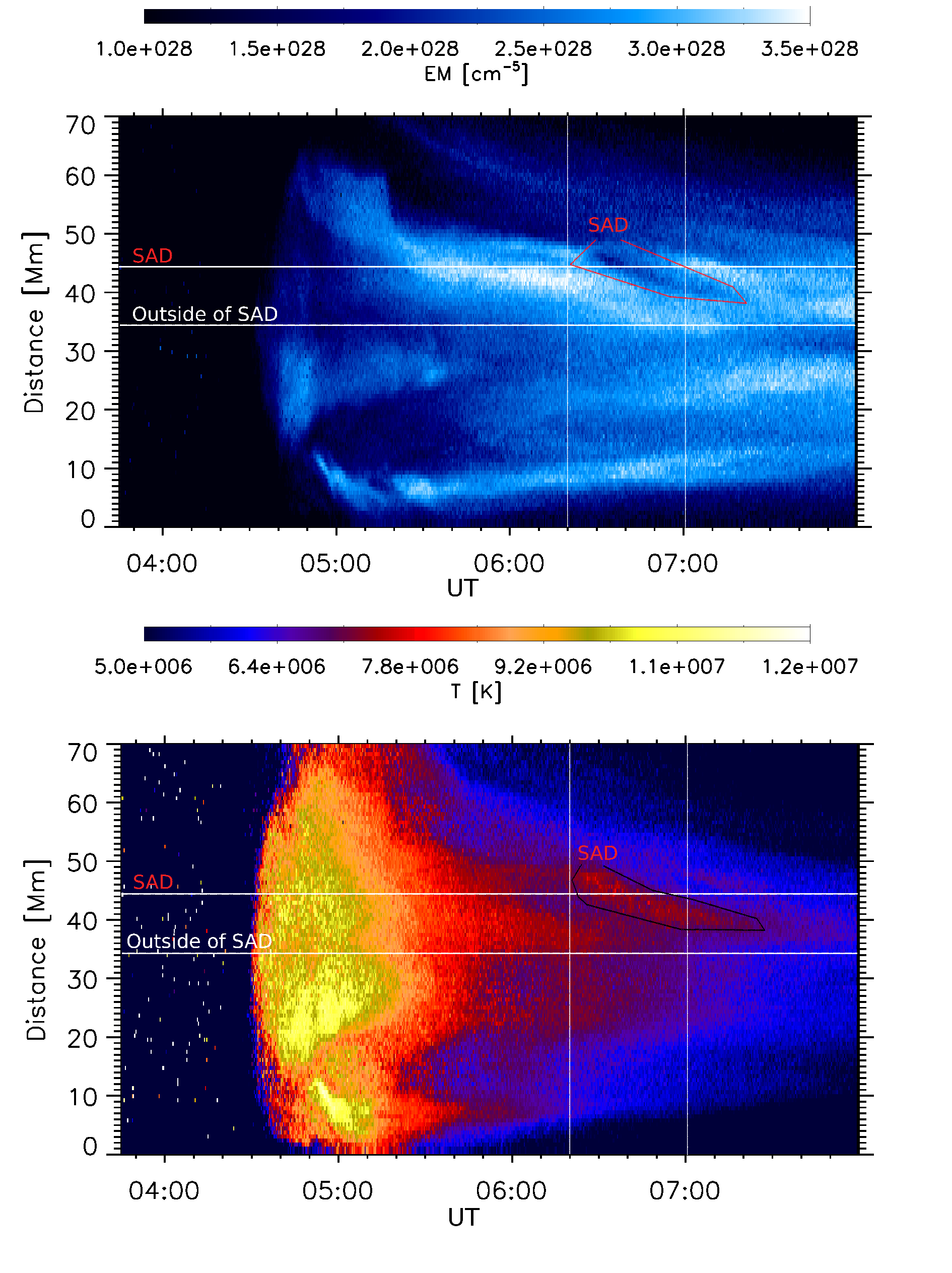}
\caption{EM and $T_{EM}$ evolution along the linear cut-a for the full time range. Two positions on the cut are selected. The first one is affected by the SAD, the second one is outside of any SADs. Two vertical lines highlight the selected time interval for the SAD analysis (cf. Fig. \ref{fig: demseriesSAD_comparison_50_62_auschnitt}).}
\label{fig: uppercut}
\end{figure}

Before dark patterns can be identified as SADs, they need to be distinguishable from the background evolution of the SAF in greater heights, where the SADs appear. Darker structures may originate from cooling, the moving branches of the  SAF, or may indeed be SADs. As is already visible from the EM and T maps, the SAF narrows and cools during the flare process. First, the temperature distribution along the linear cut-a is rather uniform (Fig. \ref{fig: uppercut}), and then a gradient forms towards the centre of the SAF.

In addition to this cooling process, there is the branched structure of the SAF, which is also visible in the stack plots of cut-a. From 04:50 UT to 05:30 UT, between $0$ and $30 \si{Mm,}$ two branches of higher EM values and lower EM values between them are accompanied by a similar pattern of higher values in the T evolution. Therefore, the lower EM values are not a SAD in this case, but originate from the position outside of the hotter and denser branches (Fig. \ref{fig: CuttingLines}).

Further dynamic changes are sometimes also faintly visible in the T maps. Although T and EM do not always correlate with each other,  in a couple of cases, patterns appearing over a limited time interval are simultaneously visible in both T and EM maps. In contrast to the branches, there are examples where low EM is accompanied by a similar pattern of higher temperature, which appear on shorter timescales. These features are in some cases SADs and are presented below.

\FloatBarrier

\subsection{SAD analysis}
\label{sec: SAD analysis}

\begin{table*}[h]
\centering   
\caption{Observed SADs and their EM and T values, before $t_{obs}$  ($EM_{1}$, $T_{1}$), during ($EM_{2}$, $T_{2}$), and after ($EM_{3}$, $T_{3}$) passage of the SAD.}
\begin{tabular}{|c|c|c|c|c|c|c|c|}
\hline
  SAD      & \begin{tabular}[c]{@{}c@{}}$t_{obs}$ \\ (UT)\end{tabular} & \begin{tabular}[c]{@{}c@{}}$EM_{1}$ \\ $[10^{28}\si{cm}^{-5}]$\end{tabular} & \begin{tabular}[c]{@{}c@{}}$T_{1}$ \\ $[\si{MK}]$\end{tabular} & \begin{tabular}[c]{@{}c@{}}$EM_{2}$ \\ $[10^{28}\si{cm}^{-5}]$\end{tabular} & \begin{tabular}[c]{@{}c@{}}$T_{2}$ \\ $[\si{MK}]$\end{tabular} & \begin{tabular}[c]{@{}c@{}}$EM_{3}$ \\ $[10^{28}\si{cm}^{-5}]$\end{tabular} & \begin{tabular}[c]{@{}c@{}}$T_{3}$ \\ $[\si{MK}]$\end{tabular} \\ \hline
1 & 06:20-07:01                                                  & 2.9                                                                                   & 6.3                                                              & 1.9                                                                                       & 6.5                                                              & 2.9                                                                                  & 6.1                                                             \\ 
2 & 06:55-07:07                                                  & 78 
 & 7                                                               & 59 
  & 7 - 6                                                          & 68                                                                                & 6                                                               \\ 
3 & 06:46 - 07:11                                                & 5                                                                                     & 6.8                                                              & 3.4                                                                                       & 7-6.2                                                            & 5                                                                                    & 6.2                                                             \\ 
4 & 04:48 - 05:13                                                & 14
 & 8.6                                                              & 11 
 & 11                                                              & 14 
 & 9.5                                                             \\ 
5 & 06:54 - 07:15                                                & 6.4                                                                                   & 7.1                                                              & 4.8                                                                                       & 6.8                                                              & 6.4
 & 6.5                                                             \\ \hline
\end{tabular}
\label{tab: SAD}
\end{table*}

One of the most pronounced EM depletions in this event reaches cut-a between 06:20 and 07:15 UT (Fig. \ref{fig: uppercut}). For the analysis, two positions on the linear cut are selected (cf. Fig. \ref{fig: uppercut}, horizontal lines). One of them is affected by the EM depletion (between 06:20 UT and 07:00 UT) and the other one is set to be close to but outside of it, as a reference for the ambient plasma conditions.

\begin{figure}[!]
\centering
\includegraphics[width=0.8\linewidth]{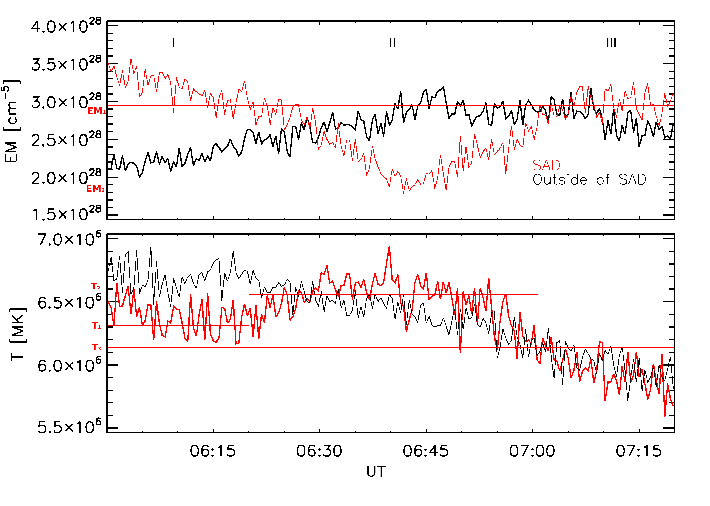}
\caption{EM and $T_\mathrm{EM}$ evolution for two positions on the linear cut-a (cf. Fig. \ref{fig: uppercut}). The red curve is data from a position passed by the SAD, while the black curve corresponds to a position outside of the SAD. Additional lines are used to highlight features in the red curves. Two vertical lines indicate the selected time interval (II), in which the temporary EM decrease occurs. Horizontal lines highlight the corresponding initial EM value ($EM_1$), minimum EM value ($EM_2$), initial temperature ($T_1$), average maximum temperature ($T_2$) during passage of the SAD, and the temperature ($T_3$) afterwards.}
\label{fig: demseriesSAD_comparison_50_62_auschnitt}
\end{figure}

The evolution of EM and T  at these two positions is plotted separately in Fig. \ref{fig: demseriesSAD_comparison_50_62_auschnitt}. The EM decreases by 45\% (corresponding to a density depletion of 21\%). In the same time range, the temperature rises by $0.3~\si{MK}$, which indicates a higher temperature along the line of sight at this position than outside of it (cf. Fig. \ref{fig: demseriesSAD_comparison_50_62_auschnitt}). The stack plots (Fig. \ref{fig: uppercut}) corroborate the finding that the feature of depleted EM is complemented by a similar feature of enhanced temperature. 

\subsubsection*{Further SADs}
As summarised in Table \ref{tab: SAD}, there are five SADs observed with durations of 15 to 40 minutes. Their detailed T and EM evolution is also displayed in the Appendix in Fig. \ref{fig: SAD2_demseries215_200_146_180_linie} - \ref{fig: SAD5_demseries215_210_150_165_linien}. While the EM depletion is in the range of 20\% to 30\%, only two SADs show an increase in temperature (SAD-1 and SAD-4). Because of the difference in the EM depletion between SAD-1 and SAD-4 (30\% and 20\%), temperature increases may happen both for stronger and weaker SADs. Based on this small group of SADs, EM depletions accompanied by temperature increases are more the exception than the rule in this event.


\subsection{Comparison with previous findings about SADs}
Recent studies found SADs reaching the apexes of flare loops, colliding with them and  causing heating in the affected regions \citep{2021Innov...200083S}. There are no SADs reaching the arcades in the event on 13 January 2015; however, their paths are directed to the hottest region within the SAF. Based on their disappearance in the diffuse plasma of the SAF, the SADs might not reach the loops, but are thwarted instead by the SAF. SADs may dissipate their energy earlier, although neither traces of heating along the SAD paths nor in front of them are observed. Therefore, the link between SADs and heating cannot be directly studied with our analysis methods. Only SAD-4 might lead to a higher temperature of the plasma behind the SAD ($T_2<T_3$).

In contrast to the predictions of models by \citet{Maglione2011} and \citet{Ccere2012}, there are no larger temperature differences between SADs and the surrounding plasma. Hence, the high temperatures of the  SAF of 10MK cannot be explained by thermal advection by SADs alone.

The absence of SADs reaching the coronal loops leads to the assumption that the SADs can be excluded as heating source for the plasma trapped in the arcades in this event. Additionally the temperature maps show that the loops and the SAF are relatively thermally isolated from each other, otherwise the great temperature difference between them of up to 3MK for extended periods of time would be balanced by heat conduction. Among other things, it is therefore of great interest to explore how the loops behave thermally in the gradual flare phase and whether or not they show an extended cooling time. An extended cooling phase would then be an indication for an additional heating process in the gradual flare phase, which is detached from heating possibly caused by SADs. Accordingly, in the following section we describe our investigation of the thermal evolution for loops and SAF separately.
\FloatBarrier 


\section{Thermal evolution in subregions}
\label{sec: Thermal Evolution in Subregions}

\begin{figure}[h]
         \centering
         \includegraphics[width=0.99\linewidth]{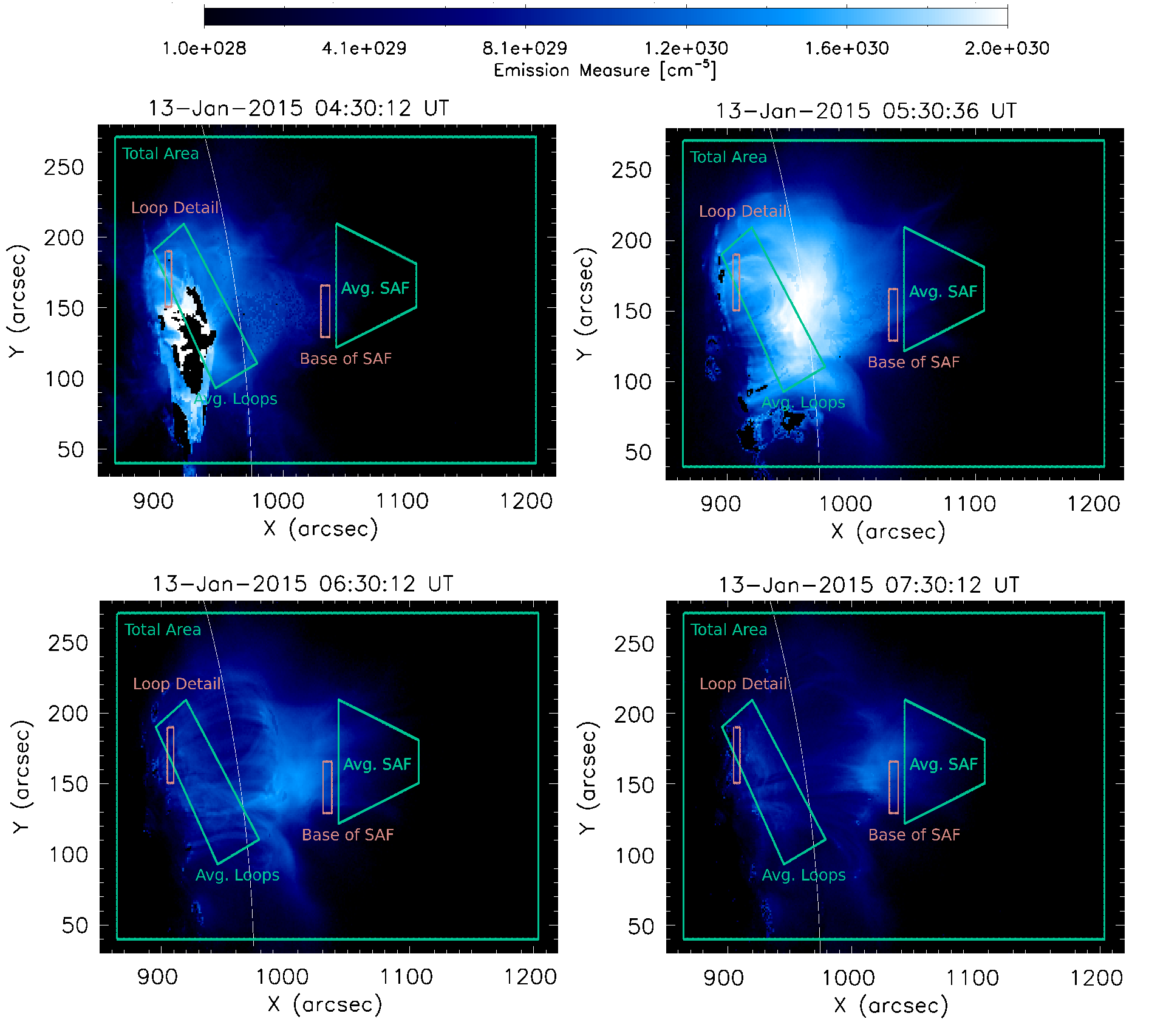}
         \caption{EM map and selected subregions. The selected subregions consist of three larger areas (green) for loops, SAF, and the total area, and two smaller regions (red) for the cooling time analysis.}
         \label{fig: subregions}
     \end{figure}

The EM and  $T_{EM}$ maps already indicate that the temperature evolves differently across the flare structures. In order to compare the thermal evolution quantitatively, the spatially resolved data can be integrated to obtain averages over certain regions of the flare. This information is extracted for two sets of subregions that are shown in Fig. \ref{fig: subregions}: 
\begin{enumerate}
    \item Averages over larger regions are selected, namely 'Avg. Loops', 'Avg. SAF' and 'Total Area'.
    \item Averages over two smaller regions, namely `Loop Detail' and `Base of SAF', are used for the calculation of cooling times in Sect. \ref{sec: Cooling Rates of Loops and SAF}.
\end{enumerate}     
For the cooling time comparison of loops and SAF (Sect. \ref{sec: Cooling Rates of Loops and SAF}), Avg. Loops, Loop Detail, and the Base of SAF -region are used.

\FloatBarrier
\subsection{DEM evolution}
\begin{figure}[h]
         \centering
         \includegraphics[width=0.98\linewidth]{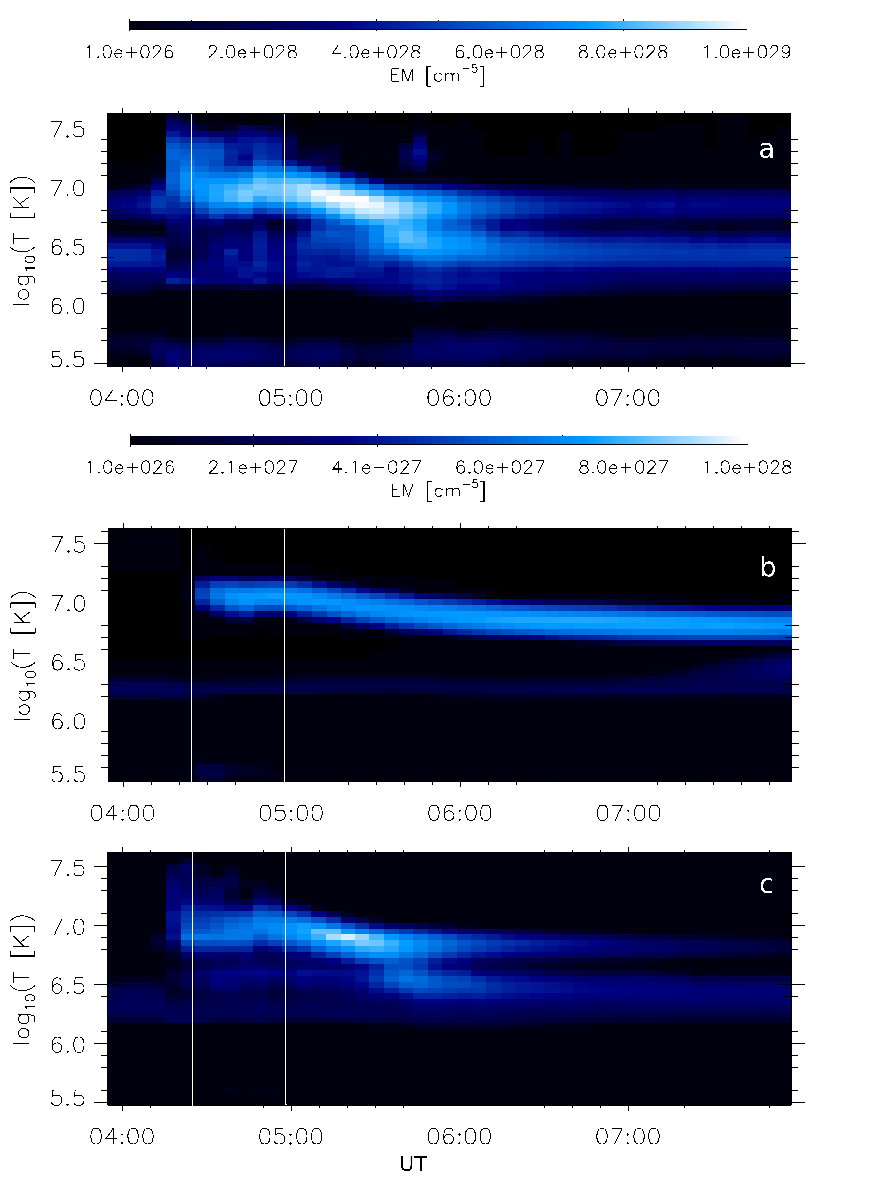}
         \caption{DEM evolution for the larger areas (green) in Fig. \ref{fig: subregions}: `Total Area' (\textit{a}), `Avg. SAF ' (\textit{b}), and the `Avg. Loops' (\textit{c}). The white lines indicate the flare peak times based on the GOES X-ray data (cf. Fig. \ref{fig: GOES_curves}).}
         \label{fig:Temperature and EM evolution for various subregions}
     \end{figure}

The time-series of reconstructed DEM cubes are used to derive the evolution of the DEM averaged over the total FOV as well as for the loop arcade and the SAF. This is shown in Fig. \ref{fig:Temperature and EM evolution for various subregions} (the values plotted correspond to the DEM integrated over the temperature bins). 

Loops and SAF contribute differently to the DEM of the Total Area. Overall, the EM is mainly distributed over two branches (Fig. \ref{fig:Temperature and EM evolution for various subregions}a). The low-temperature component ($T=1.5-3~\si{MK}$, $Log10T=6.2-6.5$) contributed by the background corona is rather constant. Additionally, there is a variable component established shortly after the flare onset that then decreases in temperature from $12~\si{MK}$ ($Log10T=7.1$) to $5~\si{MK}$ ($Log10T=6.7$) during the gradual flare phase. 
The Avg. Loops region (Fig. \ref{fig:Temperature and EM evolution for various subregions}c) shows a similar two-component evolution to the Total Area and dominates its EM, because most of the plasma is located in the loops where it is also the densest. 
In contrast, the Avg. SAF region mainly contributes a conspicuous hot component (Fig. \ref{fig:Temperature and EM evolution for various subregions}b). It forms later than the loops and remains visible in the high-temperature bins for the entire gradual flare phase, which lasts more than three hours. There is only a small background contribution in lower temperature bins, which might be a result of the location of the  subregion at greater heights.

\FloatBarrier

\section{Cooling times of loops and SAF}
\label{sec: Cooling Rates of Loops and SAF}

After having discussed the flare evolution based on AIA images as well as maps of total EM and EM-weighted temperatures derived from DEM reconstructions, in this section the measured cooling times of both flare loops and SAF are compared with the theoretical predictions of the Cargill model.

\subsection{Cargill model}
The Cargill model \citep{1995ApJ...439.1034C}, which is used in this paper as a cooling model, assumes a uniformly filled loop with a single temperature. This is supported by the idea that the plasma is frozen into the magnetic field, leading to suppressed heat conduction across the loops, while it remains unchanged along the loops. While the model was mainly applied to spatially unresolved data in previous studies (e.g. \citet{Reeves2002, Ryan2013}), here it is applied to subregions of the flare extracted from spatially resolved DEM reconstructions. These assumptions are best fulfilled in a stationary flare loop. However, the SAF is considered to consist also of loops, albeit newly reconnected ones. Thus, our quantitative analysis of the SAF cooling is restricted to the lowermost part of the SAF (close above the loop arcade) where the outflow speeds will already have decreased. Further assumptions made by the Cargill model are that:

\begin{enumerate}
\item The plasma is isotropic and isothermal.
\item There are neither flows nor heating.
\item The plasma is mono-atomic.
\item The plasma obeys the ideal gas law.
\item The plasma $\beta$ is low.
\item The conductive heat flux obeys Spitzer conductivity.
\item The radiative loss function, $P_{rad}$, is properly modelled. \citep{1978ApJ...220..643R}
\end{enumerate}

Based on these assumptions, the characteristic timescales for the conductive and radiative cooling are:

\begin{align}
\tau _\mathrm{c}  = 4 \cdot 10^{-10} \frac{nL^2}{T^{5/2}}, 
\end{align}
\begin{align}
\tau _\mathrm{r}  = 3.45 \cdot 10^{3} \frac{T^{3/2}}{n}, 
\end{align}

where $L$ is the loop half length and $n$ is the number density. The dimensions are  {centimetres (cm)}, {seconds (s)}, and Kelvin ({K)}.

Flare cooling is treated as either fully conductive or radiative at any single time. The total cooling time of the flare is given by:

\begin{align}
t_\mathrm{cgl} = \tau_{c0} \left[ \left(\frac{\tau_\mathrm{r0}}{\tau_\mathrm{c0}}\right)^{7/12} -1 \right] + \frac{2\tau_\mathrm{r0}}{3} \left(\frac{\tau_\mathrm{c0}}{\tau_\mathrm{r0}}\right)^{5/12}  \left[1-\left(\frac{T_\mathrm{L}}{T_\mathrm{0}}\right) \left(\frac{\tau_\mathrm{c0}}{\tau_\mathrm{r0}}\right)^{1/6} \right].
\end{align}

Variables with a `0' as indices are calculated at the initial time $t=0$.

Starting from a situation where $\tau_\mathrm{c0} < \tau_\mathrm{r0}$, first the flare would cool by conduction only until it reaches temperature, $T'$, which is when the two timescales are equal. From then on, the flare is assumed to cool purely radiatively to the final temperature, $T_{L}$. 
\newpage
If $\tau_\mathrm{c0} > \tau_\mathrm{r0}$ at $t=0$, then the flare is assumed to cool purely radiatively and $t_\mathrm{cgl}$ simplifies to:

\begin{align}
t_\mathrm{cgl} = \frac{2\tau_\mathrm{r0}}{3} \left(1-\frac{T_\mathrm{L}}{T_\mathrm{0}}\right).
\end{align}

The simplicity of the Cargill model, which makes it very easy to use, leads to some limitations: At any point in time the model takes only either conductive or radiative cooling into account. Also, it does not take into account enthalpy-based cooling (e.g. \citet{2010ApJ...717..163B}). Finally, it treats the flare as a single isothermal loop with constant loop half-length, although flares may consist of larger loop-systems with various physical properties.

The plasma temperatures in this flare event are high enough to neglect the enthalpy-based cooling (most significant when the temperatures of the plasma are low, i.e. below $~1-2~\si{MK}$ \citep{Ryan2013} and collisional cooling is negligible \citep{1970SoPh...15..394C}. Therefore, the Cargill model includes the most significant cooling processes in our context and is suitable for comparing cooling times derived from the DEM outputs with the predictions by this analytical cooling model. Nevertheless, this simplified model approach could be improved by using full hydrodynamic models in future projects. Spatially resolved data allow single loops and subregions to be selected. The subregions better fulfil the presented requirements for the application of the Cargill model than integrated data for the whole event,
which consists of various loop types.

\FloatBarrier
\subsection{Thermal evolution of loop arcade and SAF}

The overall thermal evolution of loops and SAF is obtained by an integration over the temperature bins (cf. Fig. \ref{fig:Temperature and EM evolution for various subregions}) and yields EM-weighted temperature and total EM curves. Here, the focus is on subregions, which qualify best for an application of the Cargill model, that is, they fulfil the assumptions stated in the previous section. Three subregions (Fig. \ref{fig: subregions}), namely  the Avg. Loops region, the Loop Detail region, and the Base of SAF region are selected. The Base of SAF region is located in the lower SAF (close to the loops), where the assumptions of a single loop length and of a stationary plasma are better fulfilled than in more dynamic regions above it.

\begin{figure}[h]
\centering
\includegraphics[width=0.95\linewidth]{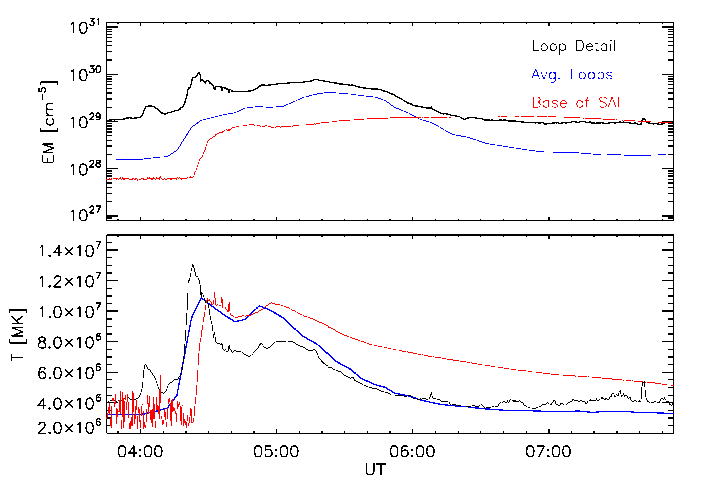}
\caption{Evolution of average EM (\textit{top}) and temperature (\textit{bottom}) for the three subregions, namely Avg. loops (blue), Base of SAF (red), and Loop Detail (black).} 
\label{fig: demseries_em_t_p}
\end{figure} 

The double peaks in the evolution of the temperatures (Fig. \ref{fig: demseries_em_t_p}) are similar to what is found from the GOES observations (Fig. \ref{fig: GOES_curves}), but there are significant differences in the maximum temperature. With the DEM routine, which takes AIA data as input, the maximum temperatures for SAF and loops remain below 14 MK, while the filter ratio method based on GOES X-ray fluxes yields temperatures of up to 20 MK.

In a comparison of thermal evolution in the SAF (Base of SAF) and the loops (Avg. Loops), there are various differences visible in both EM and T curves (Fig. \ref{fig: demseries_em_t_p}). The SAF is established later than the loops, but with a similar temperature to the loops and follows their temperature evolution up to the second peak with a delay. From there, cooling becomes dominant and the SAF cools with a much lower rate than the loops. While the average  EM decreases in the loops, it remains rather constant in the SAF. This is also visible in the DEM evolution of the Avg. Loops region (Fig. \ref{fig:Temperature and EM evolution for various subregions} b). If one assumes a constant line of sight depth, $D$, the number density of plasma, which emits in AIA EUV range, does not change significantly during the cooling phase in the SAF.

\subsection{Comparison of predicted and observed cooling times}

\begin{table*}
\caption{Values used for the Cargill model and the cooling time, $t_{cgl}$, as a result.}                 
\centering          
\begin{tabular}{|c|c|c|c|c|c|c|c|c|}     

\hline                    
  Region                                                                   & \begin{tabular}[c]{@{}c@{}}$t_\mathrm{obs}$\\ (UT)\end{tabular} & \begin{tabular}[c]{@{}c@{}}L\\ $[\si{cm}]$\end{tabular} & \begin{tabular}[c]{@{}c@{}}D\\ $[\si{cm}]$\end{tabular} & \begin{tabular}[c]{@{}c@{}}$n$\\ $[\si{cm^ {-3}}]$\end{tabular} & \begin{tabular}[c]{@{}c@{}}$T_{start}$\\ {[}$\si{MK}${]}\end{tabular} & \begin{tabular}[c]{@{}c@{}}$T_{end}$\\ {[}$\si{MK}${]}\end{tabular} & \begin{tabular}[c]{@{}c@{}}$t_\mathrm{obs}$\\ {[}$\si{min}${]}\end{tabular} & \begin{tabular}[c]{@{}c@{}}$t_\mathrm{cgl}$\\ {[}$\si{min}${]}\end{tabular} \\ \hline
\begin{tabular}[c]{@{}c@{}}Loop Detail\end{tabular} & \begin{tabular}[c]{@{}c@{}}05:08\\ -\\ 06:30\end{tabular}     & $4\cdot10^{9}$                                                                     & $2.5\cdot10^{9}$                                                     & $2\cdot10^{10}$                                                       & $8.5$                                                                 & $3.6$                                                               & $82$                                                                      & $40$                                                                  \\ \hline
Avg. Loops                                                               & \begin{tabular}[c]{@{}c@{}}05:00\\ -\\ 06:45\end{tabular}     & $5\cdot10^{9}$                                                                     & $2.5\cdot10^{9}$                                                     & $1.5 \cdot10^{10}$                                                        & $10$                                                                 & $3.5$                                                               & $105$                                                                            & $50$                                                                  \\ \hline
\begin{tabular}[c]{@{}c@{}}Base of SAF\end{tabular} & \begin{tabular}[c]{@{}c@{}}05:00\\ -\\ 07:58\end{tabular}     & $1\cdot10^{10}$                                                                    & $2.5\cdot10^{9}$                                                   & $1\cdot10^{10}$                                                       & $10$                                                                & $5.1$                                                               & $178$                                                                      & $85$\\ \hline        
   
\end{tabular}
\label{tab: Cargill parameters}       
\end{table*}

The Cargill model is applied to the thermal evolution of three subregions (defined in Sect. \ref{sec: Thermal Evolution in Subregions}). Its use requires negligible heating, and therefore we focus on the gradual flare phase. The start time is set shortly after the second temperature peak, where the cooling processes dominate the heating and the temperature decreases. Due to the different temperature peak times of the various subregions, the start times differ (Fig. \ref{fig: demseries_em_t_p}). The single hot component of the SAF supports the isothermal treatment as a good approximation (Fig. \ref{fig:Temperature and EM evolution for various subregions} b). The almost constant EM is beneficial for the use of the Cargill model, as it suggests that the amount of plasma in the loops remains relatively constant during the cooling process.

Table \ref{tab: Cargill parameters} displays the input parameters for the Cargill model, which were extracted from the AIA data (cf. Sects. \ref{sec: Observations} and \ref{sec: DEM Reconstruction Technique}), namely the loop half-lengths ($L$), calculated number densities ($N$) at the start time, the start and end temperatures ($T_\mathrm{start}$ and $T_\mathrm{end}$), and the elapsed time until the plasma has cooled down to $T_\mathrm{end}$ ($t_\mathrm{obs}$). The final column shows the cooling time $t_{cgl}$ predicted by the Cargill model, which is compared with the observed time $t_\mathrm{obs}$. 

The loop half-length for the Loop Detail is not limited by the selected region at the loop base, but is the full length of the corresponding loop and can be used with the assumption of a homogeneous loop with constant density and temperature.
The densities are based on the EM at the start time, the characteristic lengths (Table \ref{tab: Lengths}), and the assumed line-of-sight depth, $D$. The SAF is expected to be broader than its thickness \citep{2012ApJ...747L..40S}. Half of the  width of the SAF, which is $50~\si{Mm}$, is used as a line-of-sight depth for the SAF. Half of the  height of a given loop   is used as its assumed line-of-sight depth, because the loop system is relatively extended in the selected regions. While $D$ is a parameter associated with a considerable uncertainty, only its square root enters into the computation of the cooling times, and so the results are not strongly dependent on variations of this parameter. We assume 100~\si{Mm} for the loop half-length, which is of the same order of magnitude as the largest visible loops below the SAF in this event (cf. Table \ref{tab: Lengths}). Densities
are calculated with these estimated values  (eq. \ref{eq: density}). The densities in the loops ($2 \cdot10^{10} \si{cm^ {-3}}$) are found to be higher than in the SAF ($1 \cdot10^{10} \si{cm^ {-3}}$).

The cooling times predicted by the Cargill model ($t_\mathrm{cgl}$) based on the input values used here are significantly shorter than those derived from the DEM method (more than a factor of two longer, cf. Table \ref{tab: Cargill parameters}). In both subregions for the loops, the radiative cooling dominates over conductive cooling, and therefore even suppressed conduction (e.g. \citep{2016ApJ...833...76B},  \citep{2018ApJ...865...67E}) cannot explain why $t_\mathrm{obs}$ is significantly longer than $t_\mathrm{cgl}$. In the framework of the Cargill model, cooling times matching the observed ones could be obtained by using appropriately lower electron densities. However, this would imply much larger LOS depths, which are inconsistent with the AIA observations. Therefore, ongoing heating is an assumed explanation for the extended gradual phase in both loops and SAF in this flare event. This is supported by the presence of SADs, which are expected to be linked to the reconnection process and heating of coronal plasma \citep{2021Innov...200083S}. While it is not unexpected for the SAF to have longer cooling times due to the influx of energy associated with the reconnection outflow, we stress that continuous heating must also occur in the flare loops. Because of the closed magnetic field geometry of post-flare loops and the absence of SADs reaching the coronal loops, the influx of energy should not be associated with the reconnection outflow.

A two-phase heating scenario \citep{2016ApJ...820...14Q} is conceivable, because the temperature increases first impulsively and then decreases monotonically in the gradual flare phase. Low-rate heating is expected in the extended cooling phase, but may imply a significant longer heating period than the 20-30 minutes found by \citet{2016ApJ...820...14Q}.

\FloatBarrier

\section{Conclusion}
The fortunate viewing geometry of the M5 flare of 13 January 2015  allows us to gain detailed insights into the different regions within the flare and their evolution. The AIA imaging data as well as the derived DEM reconstructions indicate a clear division of the flare into a loop arcade and a SAF with very different plasma conditions.

The flare is characterised by an extended gradual phase, and both loops and SAF have a higher temperature in comparison to the pre-flare conditions for more than three hours after the flare peak. The SAF has a branched shape with alternating structures of high and low EM at greater heights above the post-flare loops. In the gradual phase, the SAF narrows and remains as a hot region directly above the loops with a decreasing temperature and EM towards its boundaries. In parallel, the loop arcade grows over time in height and becomes more ordered, as is expected, because the non-potential magnetic field energy is reduced during the flare event.

SADs were studied with respect to their temperature and density. Several previous studies found them to be predominantly cooler than the ambient SAF plasma (\citep{1999ApJ...519L..93M}, \citep{2014ApJ...786...95H}, \citep{Savage.2012}). In this event, the five observed SADs were less dense than the SAF plasma around them, which is consistent with previous studies. There is no evidence for a higher temperature component at the leading edge of the SADs (cf. \cite{2017ApJ...836...55R}), and most of them (three out of five) did not cause a noticeable change in the local temperature evolution. An exception is the largest SAD in this event (SAD-1), which had a 21\% lower density (if a constant line-of-sight depth is assumed) than the ambient plasma and is accompanied by a temperature increase of $0.3~\si{MK}$. A systematic study will be required to ascertain whether the changes in temperature and EM are correlated with the SADs, and whether there is a dependency on speed.

For flare loops and SAF, the cooling times in the gradual phase of the flare were compared with the theoretical cooling model by \citet{1995ApJ...439.1034C}. The cooling rates differ across the extended flare structure, but in all subregions, the observed cooling times are significantly longer than what is predicted by the Cargill model. Therefore, the results confirm previous studies that found the gradual phase to be longer than expected  \citep{Ryan2013}, but additionally show that this is the case for  both the loops and the SAF. Longer cooling times could be obtained from the Cargill model by adopting lower electron densities, but this would imply line-of-sight depths that are inconsistent with the AIA imaging information. Moreover, even suppressed heat conduction cannot explain the observed cooling times. The ongoing downflows in the SAF and the persistent hot component above the arcade several hours after the impulsive phase strongly suggest continuous heating during the gradual flare phase. For the SAF, this heating can straightforwardly be provided by an influx of energy associated with the reconnection outflow. However, we stress that some form of continuous heating also has to act within the loop arcade, because the closed magnetic field lines should prevent any energy input by the reconnection outflow.

Future satellite missions with stereoscopic instruments (e.g. OSCAR \citep{ refId0}) could help to better understand the three-dimensional structure of the SAF in general and SADs in particular. Further research is needed to directly link the observed plasma evolution in the loops and the SAF with the reconnection process itself. However, the application of DEM reconstruction techniques to AIA data uncovers detailed thermal evolution, which yields deep insights into the flare structure and the thermal energy distribution in its subregions.

\newpage

\begin{acknowledgements}
Part of this work was conducted in the framework of a visiting student researcher internship at the Institute of Space and Astronautical Science (ISAS) and was funded by SOKENDAI. We thank Dr. M. Cheung for his support regarding the DEM Method. Y.S. acknowledges the National Natural Science Foundation of China (grant Nos. 11820101002, U1631242, 11427803, U1731241, U1931138) and the Strategic Pioneer Program on Space Science, Chinese Academy of Sciences (grant Nos. XDA15320300, XDA15016800, XDA15320104, XDA15052200). This research is based on observations made with AIA on NASA's SDO satellite.
    
    Software: DEM sparse inversion code (\citep{2018ApJ...856L..17S}, \citep{2015ApJ...807..143C}).
\end{acknowledgements}

\FloatBarrier


%
%
\bibliographystyle{aa}
\bibliography{41868}

\begin{thebibliography}{36}
\expandafter\ifx\csname natexlab\endcsname\relax\def\natexlab#1{#1}\fi

\bibitem[{{Bian} {et~al.}(2016){Bian}, {Watters}, {Kontar}, \&
  {Emslie}}]{2016ApJ...833...76B}
{Bian}, N.~H., {Watters}, J.~M., {Kontar}, E.~P., \& {Emslie}, A.~G. 2016,
  \apj, 833, 76

\bibitem[{{Bradshaw} \& {Cargill}(2010)}]{2010ApJ...717..163B}
{Bradshaw}, S.~J. \& {Cargill}, P.~J. 2010, \apj, 717, 163

\bibitem[{{Cargill} {et~al.}(1995){Cargill}, {Mariska}, \&
  {Antiochos}}]{1995ApJ...439.1034C}
{Cargill}, P.~J., {Mariska}, J.~T., \& {Antiochos}, S.~K. 1995, \apj, 439, 1034

\bibitem[{{Carmichael}(1964)}]{1964NASSP..50..451C}
{Carmichael}, H. 1964, NASA Special Publication, 50, 451

\bibitem[{Cécere {et~al.}(2012)Cécere, Schneiter, Costa, Elaskar, \&
  Maglione}]{Ccere2012}
Cécere, M., Schneiter, M., Costa, A., Elaskar, S., \& Maglione, S. 2012,
  \apjl, 759, 79

\bibitem[{{Cheung} {et~al.}(2015){Cheung}, {Boerner}, {Schrijver}, {Testa},
  {Chen}, {Peter}, \& {Malanushenko}}]{2015ApJ...807..143C}
{Cheung}, M.~C.~M., {Boerner}, P., {Schrijver}, C.~J., {et~al.} 2015, \apj,
  807, 143

\bibitem[{{Culhane} {et~al.}(1970){Culhane}, {Vesecky}, \&
  {Phillips}}]{1970SoPh...15..394C}
{Culhane}, J.~L., {Vesecky}, J.~F., \& {Phillips}, K.~J.~H. 1970, \solphys, 15,
  394

\bibitem[{{Emslie} \& {Bian}(2018)}]{2018ApJ...865...67E}
{Emslie}, A.~G. \& {Bian}, N.~H. 2018, \apj, 865, 67

\bibitem[{{Fletcher} {et~al.}(2011){Fletcher}, {Hudson}, {Cauzzi}, {Getman},
  {Giampapa}, {Hawley}, {Heinzel}, {Johnstone}, {Kowalski}, {Osten}, \&
  {Pye}}]{2011ASPC..448..441F}
{Fletcher}, L., {Hudson}, H., {Cauzzi}, G., {et~al.} 2011, in Astronomical
  Society of the Pacific Conference Series, Vol. 448, 16th Cambridge Workshop
  on Cool Stars, Stellar Systems, and the Sun, ed. C.~{Johns-Krull}, M.~K.
  {Browning}, \& A.~A. {West}, 441

\bibitem[{{Hanneman} \& {Reeves}(2014)}]{2014ApJ...786...95H}
{Hanneman}, W.~J. \& {Reeves}, K.~K. 2014, \apj, 786, 95

\bibitem[{{Hirayama}(1974)}]{1974SoPh...34..323H}
{Hirayama}, T. 1974, \solphys, 34, 323

\bibitem[{Innes {et~al.}(2014)Innes, Guo, Bhattacharjee, Huang, \&
  Schmit}]{Innes2014}
Innes, D.~E., Guo, L.-J., Bhattacharjee, A., Huang, Y.-M., \& Schmit, D. 2014,
  \apj, 796, 27

\bibitem[{{Kopp} \& {Pneuman}(1976)}]{1976SoPh...50...85K}
{Kopp}, R.~A. \& {Pneuman}, G.~W. 1976, \solphys, 50, 85

\bibitem[{{Lemen} {et~al.}(2012){Lemen}, {Title}, {Akin}, {Boerner}, {Chou},
  {Drake}, {Duncan}, {Edwards}, {Friedlaender}, \&
  {Heyman}}]{2012SoPh..275...17L}
{Lemen}, J.~R., {Title}, A.~M., {Akin}, D.~J., {et~al.} 2012, \solphys, 275, 17

\bibitem[{{Lin} {et~al.}(2002){Lin}, {Dennis}, {Hurford}, {Smith}, {Zehnder},
  {Harvey}, {Curtis}, {Pankow}, {Turin}, {Bester}, {Csillaghy}, {Lewis},
  {Madden}, {van Beek}, {Appleby}, {Raudorf}, {McTiernan}, {Ramaty}, {Schmahl},
  {Schwartz}, {Krucker}, {Abiad}, {Quinn}, {Berg}, {Hashii}, {Sterling},
  {Jackson}, {Pratt}, {Campbell}, {Malone}, {Landis}, {Barrington-Leigh},
  {Slassi-Sennou}, {Cork}, {Clark}, {Amato}, {Orwig}, {Boyle}, {Banks},
  {Shirey}, {Tolbert}, {Zarro}, {Snow}, {Thomsen}, {Henneck}, {McHedlishvili},
  {Ming}, {Fivian}, {Jordan}, {Wanner}, {Crubb}, {Preble}, {Matranga}, {Benz},
  {Hudson}, {Canfield}, {Holman}, {Crannell}, {Kosugi}, {Emslie}, {Vilmer},
  {Brown}, {Johns-Krull}, {Aschwanden}, {Metcalf}, \&
  {Conway}}]{r2002SoPh..210....3L}
{Lin}, R.~P., {Dennis}, B.~R., {Hurford}, G.~J., {et~al.} 2002, \solphys, 210,
  3

\bibitem[{Liu(2013)}]{Liu2013}
Liu, R. 2013, Monthly Notices of the Royal Astronomical Society, 434, 1309

\bibitem[{{Maglione} {et~al.}(2011){Maglione}, {Schneiter, E. M.}, {Costa, A.},
  \& {Elaskar, S.}}]{Maglione2011}
{Maglione}, {Schneiter, E. M.}, {Costa, A.}, \& {Elaskar, S.} 2011, A\&A, 527,
  L5

\bibitem[{{McKenzie}(2000)}]{2000SoPh..195..381M}
{McKenzie}, D.~E. 2000, \solphys, 195, 381

\bibitem[{{McKenzie} \& {Hudson}(1999)}]{1999ApJ...519L..93M}
{McKenzie}, D.~E. \& {Hudson}, H.~S. 1999, \apjl, 519, L93

\bibitem[{{Moore} {et~al.}(1980){Moore}, {McKenzie}, {Svestka}, {Widing},
  {Dere}, {Antiochos}, {Dodson-Prince}, {Hiei}, {Krall}, \&
  {Krieger}}]{1980sfsl.work..341M}
{Moore}, R., {McKenzie}, D.~L., {Svestka}, Z., {et~al.} 1980, in Skylab Solar
  Workshop II, ed. P.~A. {Sturrock}, 341--409

\bibitem[{{Qiu} \& {Longcope}(2016)}]{2016ApJ...820...14Q}
{Qiu}, J. \& {Longcope}, D.~W. 2016, \apj, 820, 14

\bibitem[{{Reeves} {et~al.}(2017){Reeves}, {Freed}, {McKenzie}, \&
  {Savage}}]{2017ApJ...836...55R}
{Reeves}, K.~K., {Freed}, M.~S., {McKenzie}, D.~E., \& {Savage}, S.~L. 2017,
  \apj, 836, 55

\bibitem[{Reeves \& Warren(2002)}]{Reeves2002}
Reeves, K.~K. \& Warren, H.~P. 2002, \apj, 578, 590

\bibitem[{{Rosner} {et~al.}(1978){Rosner}, {Tucker}, \&
  {Vaiana}}]{1978ApJ...220..643R}
{Rosner}, R., {Tucker}, W.~H., \& {Vaiana}, G.~S. 1978, \apj, 220, 643

\bibitem[{Ryan {et~al.}(2013)Ryan, Chamberlin, Milligan, \&
  Gallagher}]{Ryan2013}
Ryan, D.~F., Chamberlin, P.~C., Milligan, R.~O., \& Gallagher, P.~T. 2013,
  \apj, 778, 68

\bibitem[{{Samanta} {et~al.}(2021){Samanta}, {Tian}, {Chen}, {Reeves},
  {Cheung}, {Vourlidas}, \& {Banerjee}}]{2021Innov...200083S}
{Samanta}, T., {Tian}, H., {Chen}, B., {et~al.} 2021, The Innovation, 2, 100083

\bibitem[{{Savage} {et~al.}(2010){Savage}, {McKenzie}, {Reeves}, {Forbes}, \&
  {Longcope}}]{2010AAS...21640423S}
{Savage}, S., {McKenzie}, D.~E., {Reeves}, K.~K., {Forbes}, T.~G., \&
  {Longcope}, D.~W. 2010, in American Astronomical Society Meeting Abstracts,
  Vol. 216, American Astronomical Society Meeting Abstracts \#216, 404.23

\bibitem[{Savage {et~al.}(2012)Savage, McKenzie, \& Reeves}]{Savage.2012}
Savage, S.~L., McKenzie, D.~E., \& Reeves, K.~K. 2012, \apjl, 747, L40

\bibitem[{{Savage} {et~al.}(2012){Savage}, {McKenzie}, \&
  {Reeves}}]{2012ApJ...747L..40S}
{Savage}, S.~L., {McKenzie}, D.~E., \& {Reeves}, K.~K. 2012, \apjl, 747, L40

\bibitem[{{Strugarek, Antoine} {et~al.}(2015){Strugarek, Antoine}, {Janitzek,
  Nils}, {Lee, Arrow}, {L\"oschl, Philipp}, {Seifert, Bernhard}, {Hoilijoki,
  Sanni}, {Kraaikamp, Emil}, {Mrigakshi, Alankrita Isha}, {Philippe, Thomas},
  {Spina, Sheila}, {Br\"ose, Malte}, {Massahi, Sonny}, {O\'{}Halloran, Liam},
  {Blanco, Victor Pereira}, {Stausland, Christoffer}, {Escoubet, Philippe}, \&
  {Kargl, G\"unter}}]{refId0}
{Strugarek, Antoine}, {Janitzek, Nils}, {Lee, Arrow}, {et~al.} 2015, J. Space
  Weather Space Clim., 5, A4

\bibitem[{{Sturrock} \& {Coppi}(1966)}]{1966ApJ...143....3S}
{Sturrock}, P.~A. \& {Coppi}, B. 1966, \apjl, 143, 3

\bibitem[{{Su} {et~al.}(2018){Su}, {Veronig}, {Hannah}, {Cheung}, {Dennis},
  {Holman}, {Gan}, \& {Li}}]{2018ApJ...856L..17S}
{Su}, Y., {Veronig}, A.~M., {Hannah}, I.~G., {et~al.} 2018, \apjl, 856, L17

\bibitem[{{Vaiana} {et~al.}(1973){Vaiana}, {Krieger}, \&
  {Timothy}}]{1973SoPh...32...81V}
{Vaiana}, G.~S., {Krieger}, A.~S., \& {Timothy}, A.~F. 1973, \solphys, 32, 81

\bibitem[{{{\v{S}}vestka} {et~al.}(1998){{\v{S}}vestka}, {F{\'a}rn{\'\i}k},
  {Hudson}, \& {Hick}}]{1998SoPh..182..179S}
{{\v{S}}vestka}, Z., {F{\'a}rn{\'\i}k}, F., {Hudson}, H.~S., \& {Hick}, P.
  1998, \solphys, 182, 179

\bibitem[{{Warmuth} \& {Mann}(2020)}]{2020A&A...644A.172W}
{Warmuth}, A. \& {Mann}, G. 2020, \aap, 644, A172

\bibitem[{{White} {et~al.}(2005){White}, {Thomas}, \&
  {Schwartz}}]{2005SoPh..227..231W}
{White}, S.~M., {Thomas}, R.~J., \& {Schwartz}, R.~A. 2005, \solphys, 227, 231

\end{thebibliography}

\begin{appendix}
\section{RHESSI spectrum}
\begin{figure}[h]
         \centering
         \includegraphics[width=0.9\linewidth]{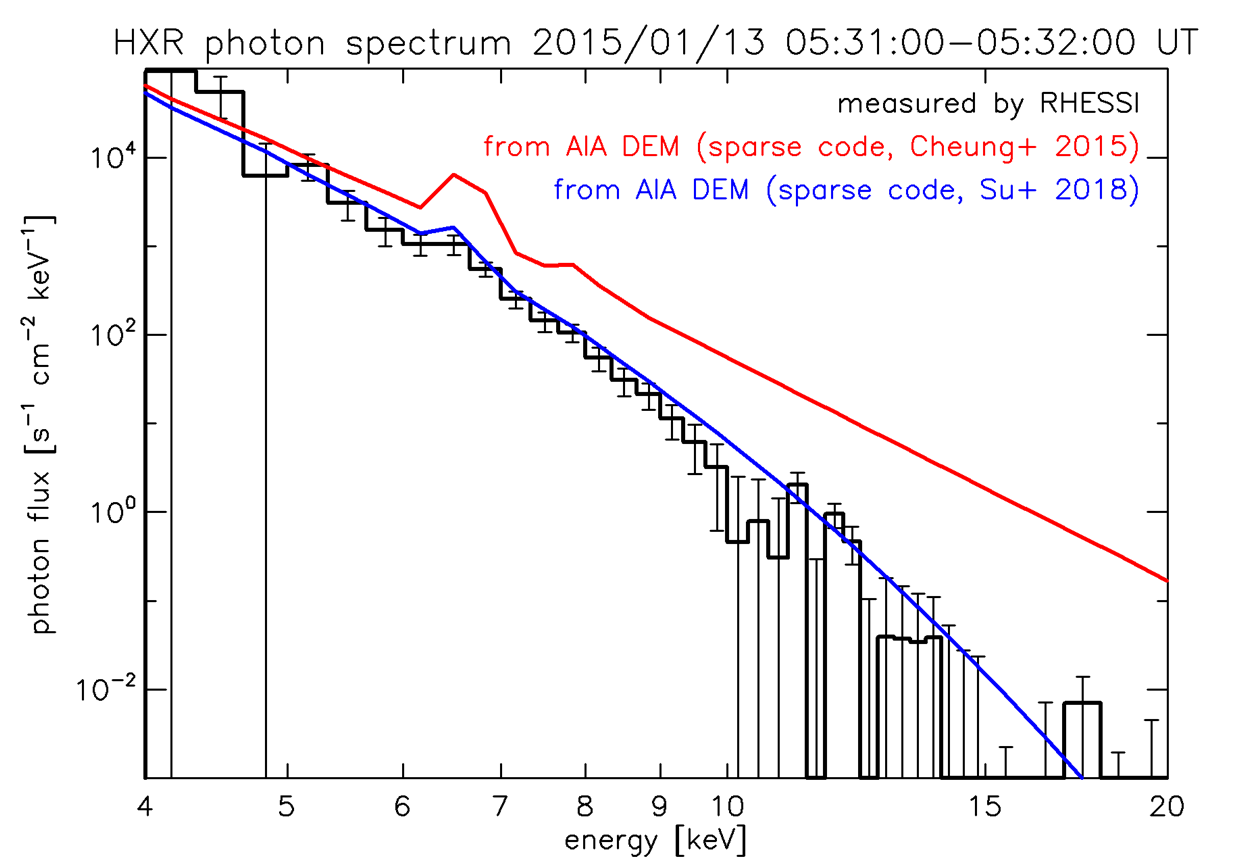}
         \caption{Comparison of a measured RHESSI X-ray photon spectrum (black) with
synthetic thermal emission spectra obtained with two DEM reconstruction techniques (red, blue). The RHESSI spectrum is generated from a background-subtracted count spectrum using detector 1, which was fitted with an isothermal component.}
         \label{fig:RHESSI_DEM_spec2}
     \end{figure}

In this section, the RHESSI X-ray photon spectrum shows the agreement between
synthetic thermal emission spectra obtained with two DEM-reconstruction techniques by \citet{2015ApJ...807..143C} and \citet{2018ApJ...856L..17S}. This is shown in Fig. \ref{fig:RHESSI_DEM_spec2} for a time-interval during the decay phase of the flare, where there are no saturation issues in the AIA data, and the X-ray emission is purely thermal.

\section{EM and T evolution of the SADs}
In this section, the evolution of EM and T  for SAD-2, SAD-3, and SAD-4 is displayed. These are discussed in Sect. \ref{sec: SAD analysis}. The curves show the evolution for certain positions on the linear cuts. The corresponding cuts are displayed in Fig. \ref{fig: CuttingLines}.

\begin{figure}[h]
\centering
\includegraphics[width=0.745\linewidth]{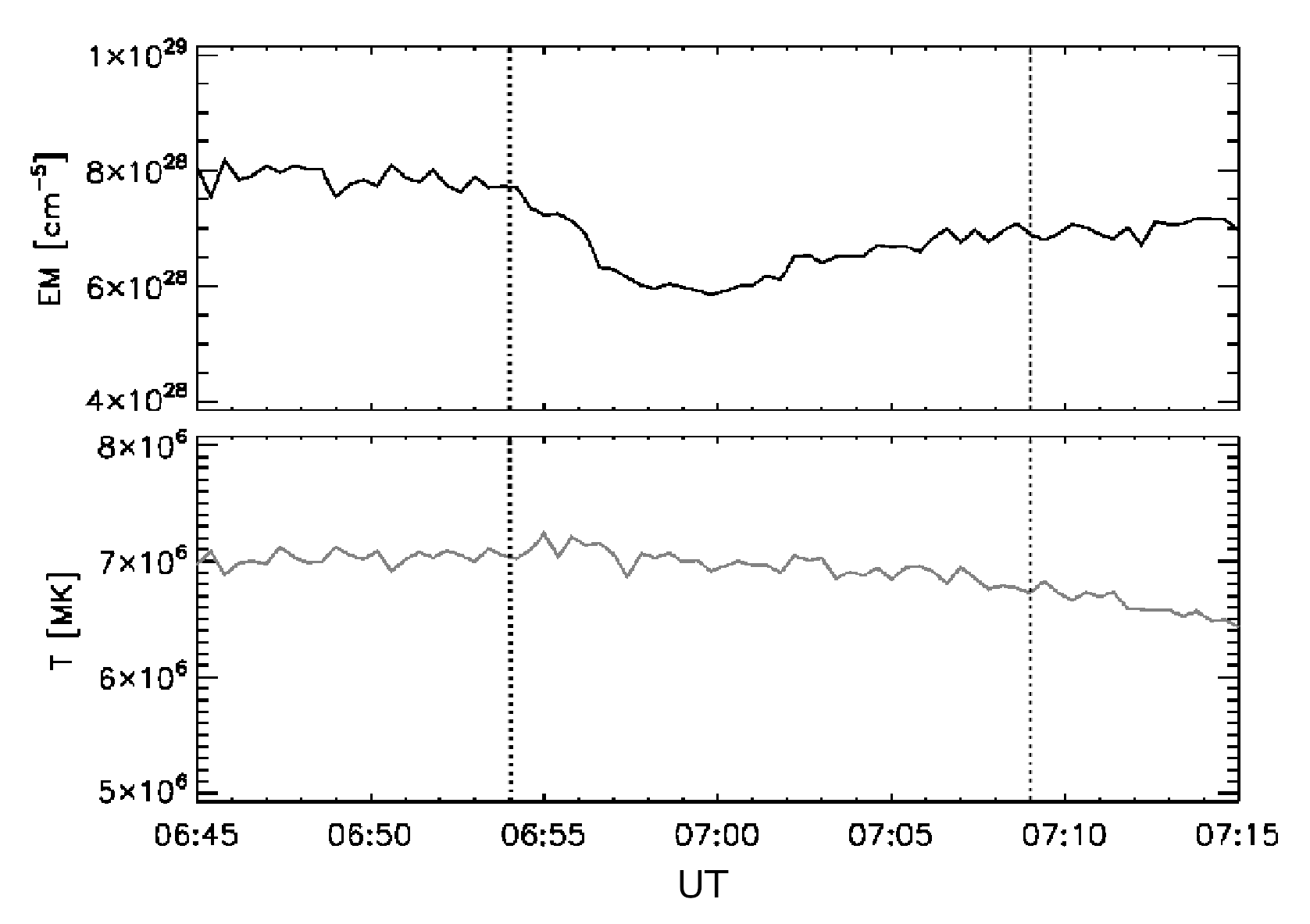}
\caption{SAD-2 on cut-b: EM and T evolution. Two dashed lines indicate the time when the SAD reaches the position on the linear cut (first) and when it has passed (second).} 
\label{fig: SAD2_demseries215_200_146_180_linie}
\end{figure}

\begin{figure}[h]
\centering
\includegraphics[width=0.745\linewidth]{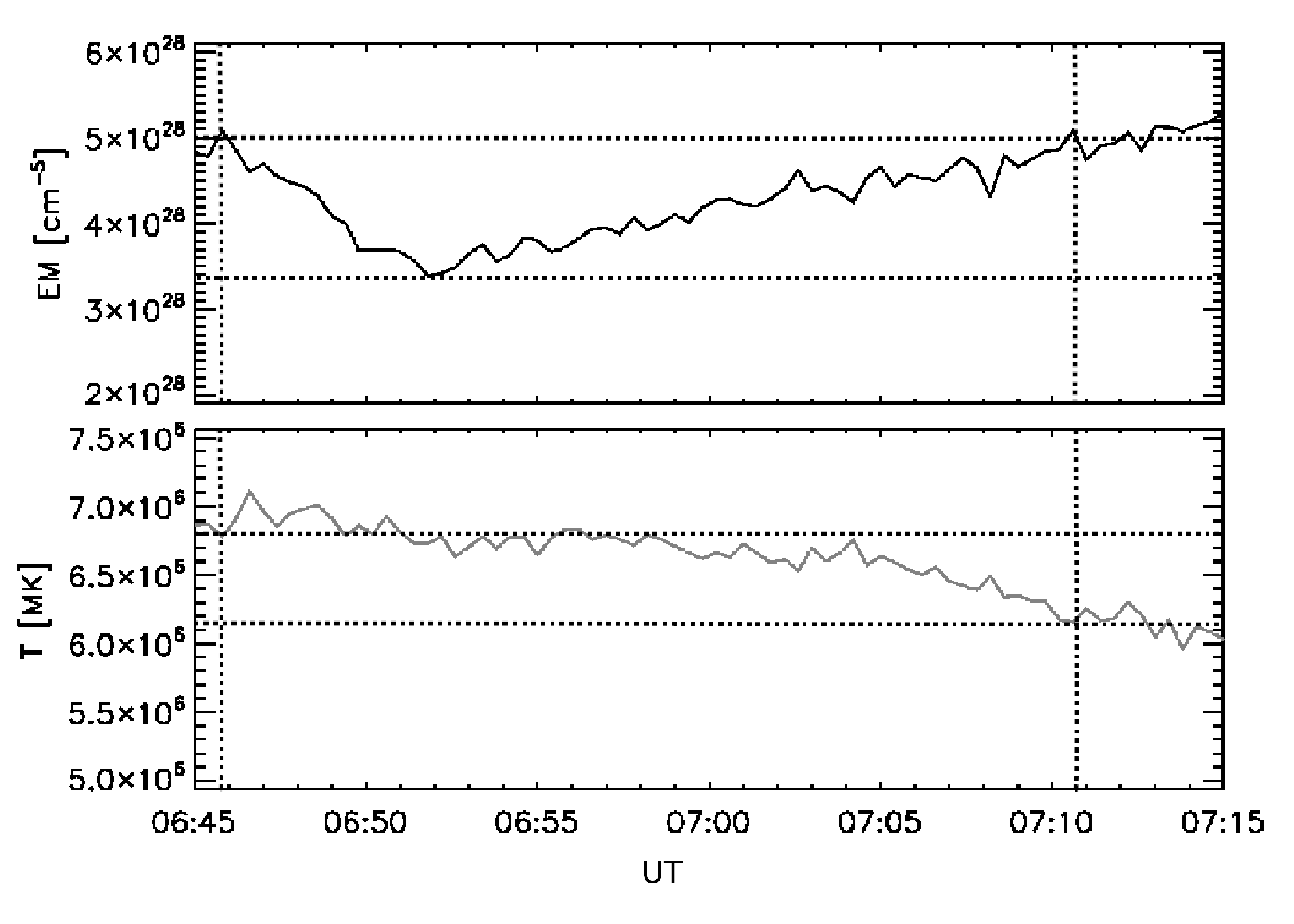}
\caption{SAD-3 on cut-b: EM and T evolution. Two dashed lines indicate the time, when the SAD reaches the position on the linear cut (first) and when it has passed (second). Horizontal dashed lines indicate the initial and minimal EM value, and the initial and final T value.} %
\label{fig: SAD3_demseries225_210_146_180_linien}
\end{figure}

\begin{figure}[h]
\centering
\includegraphics[width=0.745\linewidth]{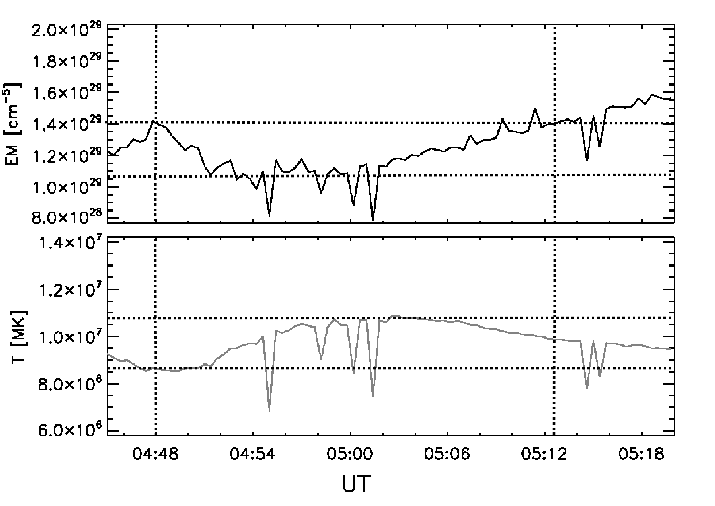}
\caption{SAD-4 on cut-c: EM and T evolution. Two dashed lines indicate the time, when the SAD reaches the position on the linear cut (first) and when it has passed (second). Horizontal dashed lines indicate the minimal and maximal EM value, and the initial and maximal T value.} 
\label{fig: SAD4_demseries193_178_106_140_linie}
\end{figure}

\begin{figure}[h]
\centering
\includegraphics[width=0.745\linewidth]{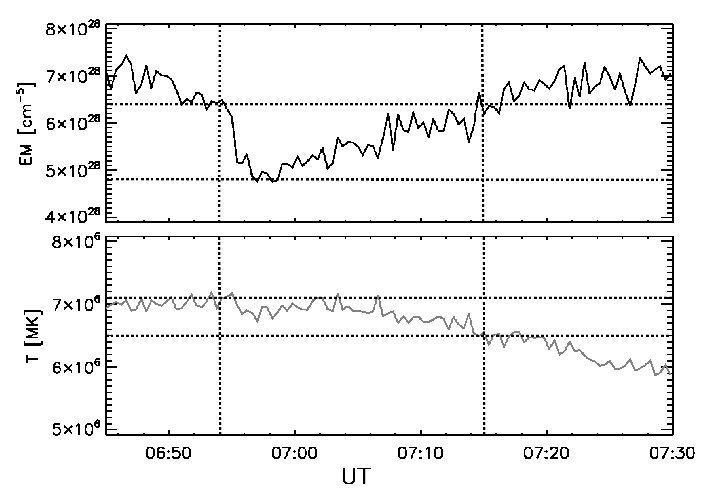}
\caption{SAD-5 on cut-b, EM and T evolution. Two dashed lines indicate the time, when the SAD reaches the position on the linear cut (first) and when it has passed (second). Horizontal dashed lines indicate the initial and minimal EM value, and the initial and final T value.} 
\label{fig: SAD5_demseries215_210_150_165_linien}
\end{figure} 

\FloatBarrier
\vspace*{5cm}
\newpage

\end{appendix}

\FloatBarrier

\end{document}